\documentclass[runningheads]{llncs}

\usepackage{rotating}  
\usepackage{bm}  
\usepackage{amsmath,amsfonts,amssymb}
\usepackage{booktabs}  
\usepackage{pdfsync}   
\usepackage[protrusion=true,expansion=true]{microtype} 
\usepackage[T1]{fontenc}
\usepackage{lmodern}

\usepackage{graphicx}
\usepackage{longtable}
\usepackage{amsfonts}
\usepackage{amssymb}
\usepackage{bm}
\usepackage[utf8]{inputenc}
\usepackage[english]{babel}
\usepackage{xstring}
\usepackage{setspace}
\usepackage{array}
\usepackage{subfig}
\usepackage{enumitem}
\usepackage{multirow}
\usepackage{algorithmic}
\usepackage{dutchcal} 			
\usepackage{ctable} 				
\usepackage{comment}
\usepackage[ruled,vlined,linesnumbered]{algorithm2e}
\usepackage{calligra}
\usepackage{balance}

\newcommand{\V}[1] {\textsf{#1}} 			

\newcommand{\Bold}[1] {\bm{#1}} 			

\newcommand{\Sensor}[1]{
    \IfEqCase{#1}{
       {IW512}{L1}
       {IW576}{L2}
       {IW544}{L3}
       {IW554}{P1}
       {IW560}{P2}
        {#1}{#1}
    }
}

\newcommand{\TE}[1]{{\scriptsize\sffamily ~#1~}} 
\newcolumntype{M}[1]{>{\centering\arraybackslash}m{#1}}
\newcolumntype{N}{@{}m{0pt}@{}}

\newcommand{\Eq}[1]{(\ref{#1})}

\newcommand{\Fig}[1]{Fig. \ref{#1}}
\newcommand{\Table}[1]{Table \ref{#1}}
\newcommand{\Sec}[1]{Section \ref{#1}}

\begin{document}

\raggedbottom
\title{Basic Computations in Fault Tree Analysis\\}
\author{Hamid Jahanian}
\institute{FS Expert (TÜV Rheinland) \#266/16-SIS\\UGL, Sydney, Australia\thanks{This article represents the author's own view, and not the position of his employer.}\\ 
\email{hamid.jahanian@ugllimited.com}\\
}

\maketitle

\begin{abstract}
Fault Tree Analysis (FTA) is a well-established method in failure analysis and is widely used in safety and reliability assessments. While FTA tools enable users to manage complex analyses effectively, they can sometimes obscure the underlying calculation processes. As a result, the soundness of FTA results often hinges on the user's expertise and familiarity with the methodology and the tool. This paper aims to explore the fundamental principles underlying both qualitative and quantitative FTA analyses, while addressing broader conceptual considerations such as coherence and consensus. By developing a deeper understanding of these concepts, engineers can improve their ability to interpret, verify, and make informed use of the outputs generated by FTA tools. This paper does not propose a novel concept in FTA but aims to compile and present a concise overview of the fundamental computations in FTA.

\end{abstract}

\section{Introduction}\label{Sec_Introduction}

Fault Tree Analysis (FTA) was first introduced in 1961 to analyze potential faults in a missile launch control system \cite{Ref_390}. Over the past six decades, numerous extensions and variations of the method have been developed to address a wide range of problems \cite{Ref_138,Ref_145}. In this article, we review the standard FTA methodology, aiming to provide a concise yet comprehensive guide to FTA computations for those interested in its practical implementation.

The article begins with a brief introduction to FTA in \Sec{Sec_FTAQQ}, followed by a discussion on the probabilistic modeling of basic events in \Sec{Sec_FTABEs}. Sections \ref{Sec_FTAMCSs} and \ref{Sec_QuantitativeFTA} delve into the qualitative and quantitative aspects of FTA computations, respectively. Finally, \Sec{Sec_BeyondCalc} addresses some of the challenges associated with FTA, including NOT gates, non-coherence, Consensus Theorem, and considerations for selecting the most appropriate models and tools. Each section outlines key concepts, complemented by illustrative examples to enhance understanding.\footnote{Sections \ref{Sec_FTAQQ} to \ref{Sec_QuantitativeFTA} of this paper have been adapted from Appendix B of the author's PhD thesis, titled ``Failure Mode Reasoning in Safety-Critical Programs'' \cite{Ref_360}.}

This paper draws from established textbooks, academic articles, and the author's own practical experience. Readers are encouraged to refer to the original sources cited for a more in-depth exploration of the topics discussed.
 
\section{Basics of FTA}\label{Sec_FTAQQ}

\begin{figure}[!h]  
\begin{center}
\includegraphics[scale=0.6]{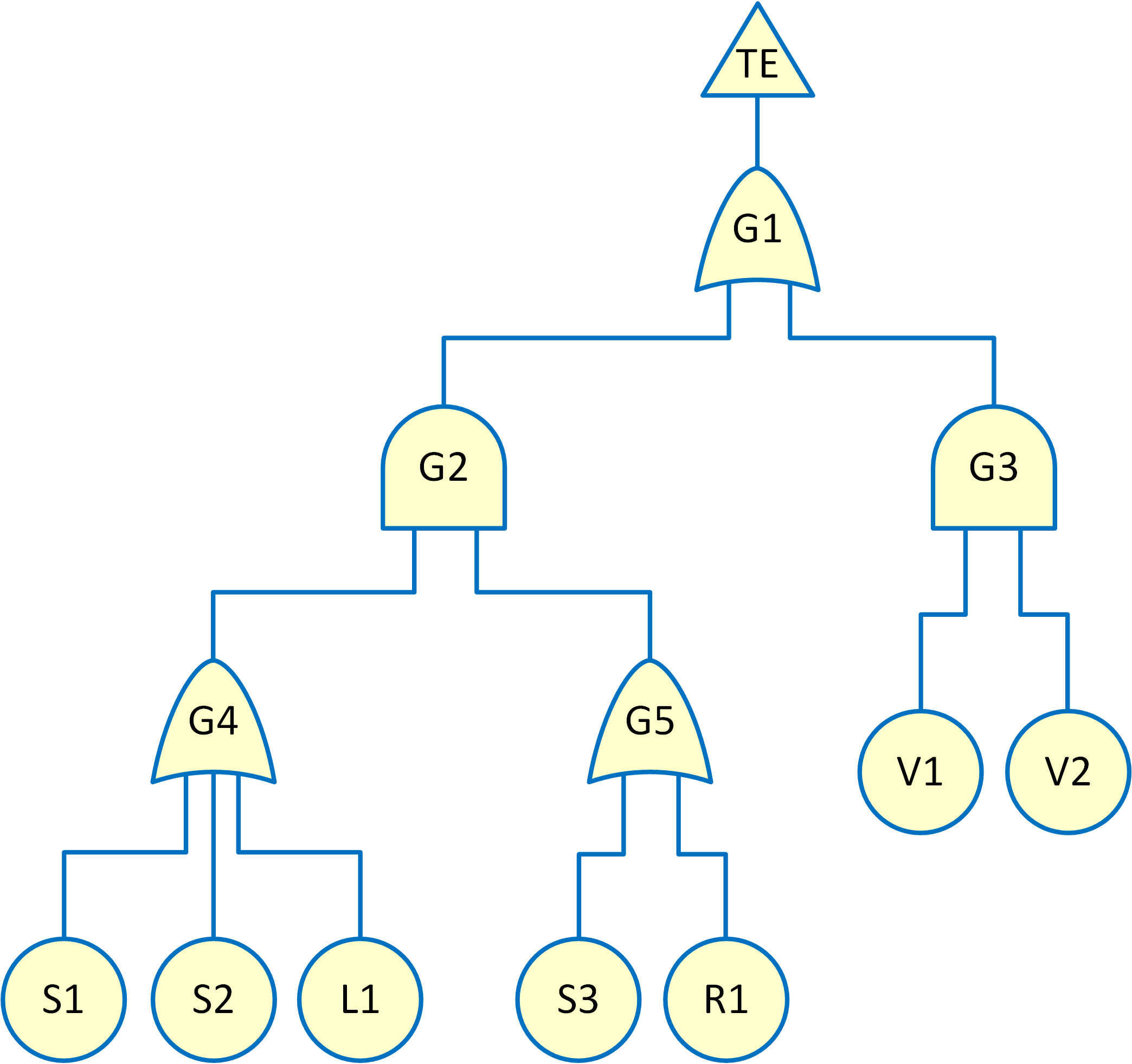}
\end{center}
\caption{Example fault tree}
\label{Fig_FT_Ex1}
\end{figure}

A fault tree is a graphical representation of failure in a system. Fault trees comprises events and gates. Events are categorized into basic events, which are not the effect of other events in the model; intermediate events, which are caused by a combination of other events; and a top event, which is the overall failure at the system level. \Fig{Fig_FT_Ex1} shows an example fault tree, comprising seven basic events ($S1, S2, L1, S3, R1, V1$, and $V2$), three OR gates ($G1, G4$, and $G5$), and two AND gates ($G2$ and $G3$). The top event in this tree is the output of $G1$. 

The output of an AND gate occurs if all of its inputs occur. For an OR gate, the occurrence of one input leads to the occurrence of the output. More complex types of gates are typically used in FTA. Readers can refer to sources such as \cite{Ref_182} and \cite{Ref_168} for details.

The graphical representation of fault trees helps analysts visually navigate through the trees. Fault trees can also be expressed in mathematical terms, which enables automated tools to \emph{solve} them. One main task in solving a fault tree is to simplify it into its Minimal Cut Sets (MCSs). An MCS is the smallest set of basic events that can together result in the occurrence of the top event. This part of FTA is known as qualitative analysis.

Fault trees are also used for quantitative analyses, where an estimation of the probability of system failure is sought. In simple terms, such calculations are based on the basic calculations in intersections and unions.\footnote{Some sources use the terms ``intersection'' and ``union,'' and the symbols $\cap$ and $\cup$. However, this should not be confused with the intersection and union of sets. When speaking of occurrence of events, $\cap$ and $\cup$ represent \emph{logical} intersection and union (i.e., conjunction and disjunction).}
For an AND gate expressed by $C=A \wedge B$, the probability of $C$ occurring can be calculated by the following \cite{Ref_168}:\footnote{We use the symbols $\wedge$, $\vee$, and $\neg$ for logical AND, OR, and negation, respectively.}
\begin{equation}\label{Eq_ANDPr}
Pr(C)=Pr(A) \cdot Pr(B|A)=Pr(B) \cdot Pr(A|B)
\end{equation}
where $Pr(X)$ means the probability of event $X$ occurring, and $A|B$ means event $A$ occurring, given that event $B$ has already occurred. In cases where events $A$ and $B$ are independent, $Pr(B|A)=Pr(B)$ and $Pr(A|B)=Pr(A)$, and thus \Eq{Eq_ANDPr} can be simplified to the following:
\begin{equation}\label{Eq_IndANDPr}
Pr(C)=Pr(A) \cdot Pr(B)
\end{equation}

In comparison with AND gates, which represent the conjunction of events, an OR gate formulates their disjunctions. For the OR gate $D=A \vee B$, the probability of $D$ occurring can be calculated by the following \cite{Ref_168}:
\begin{align}\label{Eq_ORPr}
Pr(D)&=Pr(A) + Pr(B) - Pr(A \wedge B) \nonumber \\ &= Pr(A) + Pr(B) - Pr(A) \cdot Pr(B|A) \nonumber \\ &= Pr(A) + Pr(B) - Pr(B) \cdot Pr(A|B)
\end{align}
If $A$ and $B$ are mutually exclusive, then $Pr(A \wedge B)=0$ and thus: 
\begin{align}
Pr(D)=Pr(A) + Pr(B) 
\end{align}
If $A$ and $B$ are independent, then $Pr(B|A)=Pr(B)$ and $Pr(A|B)=Pr(A)$, and thus:
\begin{align}\label{Eq_ConsORPr}
Pr(D)&=Pr(A) + Pr(B) - Pr(A) \cdot Pr(B) \nonumber \\
&=1-(1-Pr(A))(1-Pr(B))
\end{align}

The results of a quantitative FTA heavily rely on the failure models of the individual basic events, which are defined independently from the architecture of the tree. In the next section we explain the reliability concepts for modeling basic events. 

\section{Modeling the basic events}\label{Sec_FTABEs}

FTA tools commonly use unavailability and failure frequency measures for the quantitative evaluation of fault trees. For a repairable item,\footnote{A repairable item is one that can be repaired and restored to its functional state once found faulty.} unavailability $q(t)$ is the probability that the item is not in a functioning state at time $t$ \cite{Ref_204} and failure frequency $w(t)$ the probability that the item fails per time unit at time $t$, also known as the ``unconditional'' failure intensity \cite{Ref_207}. In general, the relation between the two is defined by the following:  
\begin{equation}\label{Eq_WLQ}
w(t)=\lambda(t)(1-q(t))
\end{equation}
in which $\lambda(t)$ is the ``conditional'' failure intensity \cite{Ref_207}, commonly known as failure rate. 

FTA tools generally use time-independent calculations in which the constant approximations of $\lambda$, $w$, and $q$ are used. For a basic event, unavailability $q$ is determined based on the failure characteristics of the event, including, for instance, the failure rate of a corresponding component. 

Commonly used failure models for basic events include ``Fixed Probability,'' ``Failure-Repair Rate,'' and ``Dormant.''\footnote{Readers may find FTA tools that use different sets of names for their models.} In what follows, we review the formulation of $w$ and $q$ for these three types. 

The Fixed Probability model is used when the occurrence of the basic event is expressed independently from the repair process. For a basic event with a constant (fixed) probability of failure $p$, the unavailability will be as follows:
\begin{align}\label{Eq_Fixedq}
q=p
\end{align}

The Failure-Repair Rate model is suitable for repairable elements. These are the components of which the occurrence of faults is detected, and the detection of faults leads to the repair and restoration of elements. Thus, the only time that the component is unavailable is the time that it is under repair. This time interval is known as Mean Time to Restore (MTTR). The unavailability of such components is modeled by the following \cite{Ref_207}:
\begin{equation}\label{Eq_Rateq}
q(t)=\dfrac{\lambda}{(\lambda+\mu)}(1-e^{-(\lambda+\mu)t})
\end{equation}    
Here, $\lambda$ is the failure rate and $\mu=1/MTTR$ is the restoration (repair) rate. These rates are often expressed in a \textit{per hour} ($\V{ph}$) unit.\footnote{Throughout this paper, we use $\V{ph}$, $\V{h}$, $\V{py}$, and $\V{y}$ for ``per hour,'' ``hours,'' ``per year,'' and ``years'' units, respectively.} To calculate $q$ as an independent measure from time, $t$ is often assigned the maximum value relevant to the assessment period. For an assessment period of one year, for instance, $t$ will be 8,760 hours, which equates one year. Failure-Repair Rate models can also be used for non-repairable items by assigning $\mu=\V{0.0}$.  

The Dormant type is used when a basic event depicts an undetected fault of a component that undergoes periodic proof testing. Dangerous Undetected failure in demand-mode Safety Instrumented Functions (SIFs) is an example in which Dormant models are best suited. In effect, Dormant components exhibit characteristics similar to non-repairable components during periods between two proof tests. Some FTA tools \cite{Ref_289,Ref_290} use the following to estimate the unavailability of such components:
\begin{equation}\label{Eq_Dormantq}
q=\dfrac{\lambda\tau-(1-e^{-\lambda\tau})+\lambda\cdot MTTR \cdot (1-e^{-\lambda \tau})}{\lambda\tau+\lambda \cdot MTTR \cdot (1-e^{-\lambda \tau})}
\end{equation}    

Here we justify this formula. Consider a component with failure rate $\lambda$ and proof test period $\tau$. At $t=0$, the component is fault-free. The probability that the component is faulty at time $t<\tau$ is $F(t)=1-e^{-\lambda t}$ \cite{Ref_204}. Therefore, the probability that the component is found faulty at the end of a $\tau$ cycle, and thus requires a repair, is $F(\tau)=1-e^{-\lambda \tau}$. Note that MTTR is the mean time required to repair and restore the equipment, but this time is only spent if the component is found faulty. The mean time that is actually spent on repair and restoration at the end of each test cycle is $T_R=(1-e^{-\lambda \tau}) MTTR$, which means that the time between the beginning of two subsequent test cycles will actually be $T=\tau+T_R$, instead of the predefined $\tau$, with $T$ being the following:
\begin{align}\label{Eq_Dormantq1}
T=\tau+T_R=\tau+(1-e^{-\lambda \tau}) MTTR
\end{align}

Now, let $T_U$ be the mean time during which the component is unavailable within one $T$ interval. Therefore, the unavailability of the component can be measured by $q=T_U/T$. The mean time $T_U$, itself, comprises two parts: the mean time within a $\tau$ cycle during which the component is faulty and the fault is undetected, namely $T_D$; and the mean time when the component undergoes repair and restoration (i.e., $T_R$). $T_D$ can be calculated by $T_D=\int_{0}^{\tau}F(t) dt$ \cite{Ref_204}. Therefore, $T_U$ can be calculated as follows:
\begin{align}\label{Eq_Dormantq2}
T_U=T_D+T_R&=\dfrac{\lambda \tau-(1-e^{-\lambda \tau})}{\lambda} + (1-e^{-\lambda \tau}) MTTR
\end{align}

Recall that $q=T_U/T$. By using \Eq{Eq_Dormantq2} and \Eq{Eq_Dormantq1}, $q=T_U/T$ yields \Eq{Eq_Dormantq}. 

In scenarios in which $MTTR\ll \tau$, \Eq{Eq_Dormantq} can be simplified to the following:
\begin{equation}
q=1-\dfrac{1-e^{-\lambda\tau}}{\lambda\tau}
\end{equation} 

Frequency of occurrence is another measure for expressing the likelihood of the basic events. Basic events with fixed probability values cannot be expressed in frequency form. Fixed Probability is often used to model enabling conditions, such as a proportion of time or a condition under which the occurrence of other basic events can lead to the top event. For the Failure-Repair Rate and Dormant models, the frequency of the basic event occurring will be as follows:
\begin{equation}\label{Eq_BEFreq}
w=\lambda(1-q)
\end{equation}

The modeling concepts we discussed here are common in various reliability analysis methods, including FTA, Event Tree Analysis (ETA), and Reliability Block Diagrams (RBD). The same concepts were also used in Failure Mode reasoning (FMR) for calculating the likelihood of input failure modes. 

\subsection{Example}\label{Sec_BE_Example}

The fault tree in \Fig{Fig_FT_Ex1} models the failure of a SIF comprising sensors $S1, S2$, and $S3$, Logic Solver $L1$, relay $R1$ and valves $V1$ and $V2$. In this model, the SIF fails if both paths for detecting hazard fail -- where $S1, S2$, and $L$ constitute one path and $S3$ and $R$ another -- or if both valves ($V1$ and $V2$) fail to act to contain the hazard. Below, we model these devices based on the information given for each type of component. We suppose $MTTR=\V{8 hours}$ for all components. 

Suppose we know that sensors $S1, S2$, and $S3$ are identical with a failure rate of $\lambda_S=\V{5.0E-8 ph}$ and a proof test interval of four years ($\V{35,040 hours}$). Based on the information given, the Dormant model will be best suitable for these devices, as they have undetected failure modes and undergo regular proof tests. Thus, by using \Eq{Eq_Dormantq} and \Eq{Eq_BEFreq}, we can calculate $q_S=\V{8.759E-4}$ and $w_S=\V{4.996E-8 ph}$. 

Next, suppose the Logic Solver $L1$ is not subject to proof testing and repair and has a constant PFD of $\V{5.6E-4}$. By using the Fixed Probability model and \Eq{Eq_Fixedq} we will have $q_L=\V{5.600E-4}$. Using a Fixed Probability model also means $w_L=\V{0.0}$.

Let us suppose that the relay $R1$ is a simple device with no undetected failure modes and thus no requirements for proof testing, and its failure rate is $\lambda_R=\V{6.0E-8 ph}$. Failure-Repair Rate will be suitable for this device. By using \Eq{Eq_Rateq} and \Eq{Eq_BEFreq} where $t=\V{35,040 hours}$, we will have $q_R=\V{4.800E-7}$ and $w_R=\V{6.000E-8 ph}$.

Finally, we suppose that the valves $V1$ and $V2$ are identical with a failure rate of $\lambda_V=\V{1.5E-6 ph}$ and a proof test interval of one year ($\V{8,760 hours}$). Similar to the sensors, we use the Dormant type to model the valves, which results in $q_V=\V{6.553E-3}$ and $w_V=\V{1.490E-6 ph}$.

\section{Qualitative FTA}\label{Sec_FTAMCSs}

In FTA, a ``cut set'' is a set of events that if occur together, can result in the occurrence of the top event. Accordingly, an MCS is a ``minimal'' cut set, meaning that the events included in the MCS are independent, and all of them are necessary for the occurrence of the top event. 

A fault tree can be simplified to its MCSs. Among methods and algorithms designed for computing MCSs, MOCUS \cite{Ref_178} and MICSUP \cite{Ref_179} are well known. The former method takes a top-down approach and starts at the top of the tree, while the latter method is a bottom-up approach and starts at the basic events. 

The MOCUS algorithm uses an expandable table that starts with one cell, the top event, and finishes with the list of all possible cut sets \cite{Ref_138}. A filtering process is conducted at the end to remove the redundant cut sets and shrink the table down to the list of MCSs. At each step of the algorithm, an output of a gate is replaced by its inputs. For AND gates, this replacement is done at the spot by replacing the output variable with the combination of input variables. For OR gates, new rows are added to cater for all possible combinations. 

\subsection{Example}\label{Sec_MCS_Example}

For the fault tree we showed in \Fig{Fig_FT_Ex1}, the MOCUS table starts at $G1$, which produces the top event. Since $G1$ is an OR gate, its inputs will be listed in separate rows:
\begin{longtable}[!h]{|M{3.0cm}|M{3.0cm}|}
\hline \TE{Iteration 1:} 
& \TE{$G2$} \\ [-0.1em]
& \TE{$G3$} \\ \hline
\end{longtable}
We now expand on $G2$, which is an AND gate and needs no new lines to be added:
\begin{longtable}[!h]{|M{3.0cm}|M{3.0cm}|}
\hline \TE{Iteration 2:} 
& \TE{$G4, G5$} \\ [-0.1em]
& \TE{$G3$} \\ \hline
\end{longtable}
The process continues until all the gates are expanded with respect to basic events:
\begin{longtable}[!h]{|M{3.0cm}|M{3.0cm}|}
\hline \TE{Iteration 3:} 
& \TE{$G4, G5$} \\ [-0.1em]
& \TE{$V1, V2$} \\ \hline
\hline \TE{Iteration 4:} 
& \TE{$S1, G5$} \\ [-0.1em]
& \TE{$S2, G5$} \\ [-0.1em]
& \TE{$L1, G5$} \\ [-0.1em]
& \TE{$V1, V2$} \\ \hline
\hline \TE{Iteration 5:} 
& \TE{$S1, S3$} \\ [-0.1em]
& \TE{$S1, R1$} \\ [-0.1em]
& \TE{$S2, S3$} \\ [-0.1em]
& \TE{$S2, R1$} \\ [-0.1em]
& \TE{$L1, S3$} \\ [-0.1em]
& \TE{$L1, R1$} \\ [-0.1em]
& \TE{$V1, V2$} \\ \hline
\end{longtable}
The final entries in this example (i.e., in Iteration 5) are already independent and require no further filtering. Therefore, the MCSs are as follows: 

\begin{align}
&\{S1, S3\}, \{S1, R1\}, \{S2, S3\}, \{S2, R1\}, \{L1, S3\}, \{L1, R1\}, \{V1, V2\} 
\end{align}
Accordingly, the occurrence of the top event $TE$ can be expressed in a logical form as follows:
\begin{align}
TE=&(S1 \wedge S3) \vee (S1 \wedge R1) \vee (S2 \wedge S3) \vee (S2 \wedge R1) \vee \nonumber \\
&(L1 \wedge S3) \vee (L1 \wedge R1) \vee (V1 \wedge V2)
\end{align}

\section{Quantitative FTA}\label{Sec_QuantitativeFTA}

In a fault tree, the likelihood of the top event and MCSs can be calculated based on the likelihood of their constituting basic events. Let $E=\{e_i ~|~ 1 \leq i \leq n\}$ be an MCS with $n$ basic events. Let $q_i$ be the probability of occurrence of basic event $e_i$, and $Q_{E}$ the unavailability measure for $E$. Since the basic events in an MCS are independent, we can use \Eq{Eq_IndANDPr} to say the following:
\begin{align}
&Q_{E}=\prod_{i=1}^{n}q_i \label{Eq_MCSQ}
\end{align}
Given that $0 \leq q_i \leq 1$, the upper bound of $Q_{E}$ will be as follows:
\begin{align}
&Q_{E} \leq min\{q_i ~|~ 1 \leq i \leq n\} \leq q_i
\end{align}

Let $w_i$ be the frequency of occurrence of basic event $e_i$ and $W_{E}$ the overall frequency of occurrence of $E$. Given that MCS is a conjunctive form, an MCS occurs only when all its constituting basic events occur. This requires the last basic event to occur when all the other basic events are already in the failed state. Therefore:
\begin{align}
&W_{E}=\sum_{i=1}^{n}w_i\prod_{\substack{k=1 \\  k\neq i}}^{n}q_k=Q_{E}\sum_{i=1}^{n}\dfrac{w_i}{q_i} \label{Eq_MCSW} 
\end{align}

Now, let $TE$ be the top event of a fault tree with $m$ MCSs where an MCS is depicted by $E_i=\{e_{ij} ~|~ 1 \leq j \leq n\}$, with $1 \leq i \leq m$. Let $Q_{i}$ and $W_{i}$ represent the unavailability and frequency of $E_i$. The unavailability of $TE$ can be calculated as follows \cite{Ref_204,Ref_207}:\footnote{Recall that $\cap$ and $\cup$ represent \emph{logical} intersection and union.}
\begin{align}\label{Eq_AppB_3}
Q_{TE} &= Pr\left(\bigcup_{i=1}^{m}E_i\right) \nonumber\\
             &=\sum_{i=1}^{m}Pr(E_i)-\sum_{i<j}Pr(E_i \cap E_j)+\sum_{i<j<k}Pr(E_i \cap E_j \cap E_k)- ... \nonumber \\
             &+(-1)^{m+1}Pr(\bigcap_{i=1}^{m}E_i) \nonumber \\
		&=\sum_{i=1}^{m}Q_i-\sum_{i<j}Pr(E_i \cap E_j)+\sum_{i<j<k}Pr(E_i \cap E_j \cap E_k)- ... \nonumber\\
             &+(-1)^{m+1}Pr(\bigcap_{i=1}^{m}E_i)
\end{align} 

Relation \Eq{Eq_AppB_3} is known as the Inclusion-Exclusion (IE) formula and is used for estimating the unavailability of the top event in a fault tree. While the method provides the most accurate estimation, it is a hard-to-implement process for solving large-scale fault trees in computer-based tools. Therefore, a conservative yet realistic approximation of IE formula would be preferable. 

We know that for any given $m^{\prime}, m^{\prime\prime} \leq m$, if $m^{\prime} \leq m^{\prime\prime}$ then $Pr(\bigcap_{i=1}^{m^{\prime}}E_i) \geq Pr(\bigcap_{i=1}^{m^{\prime\prime}}E_i)$. Therefore, in \Eq{Eq_AppB_3}:
\begin{align}
&\sum_{i=1}^{m}Pr(E_i) \geq \sum_{i<j}Pr(E_i \cap E_j) \geq \sum_{i<j<k}Pr(E_i \cap E_j \cap E_k) \geq ... 
\end{align} 
which implies that:
\begin{equation}\label{Eq_AppB_4}
Q_{TE} \leq \sum_{i=1}^{m}Q_i
\end{equation} 

Inequality \Eq{Eq_AppB_4} depicts the most conservative upper bound for $Q_{TE}$. This limit, also known as Rare-Event approximation \cite{Ref_176}, is easy to calculate but can result in overestimation. 

Esary and Proschan proved the lower and upper bounds for the reliability of a system \cite{Ref_201,Ref_202}. The lower bound of reliability is the upper bound of unavailability. Therefore: 
\begin{equation}\label{Eq_AppB_5}
Q_{TE} \leq 1-\prod_{i=1}^{m}(1-Q_i)
\end{equation} 

It can be shown that the Esary-Proschon (EP) upper bound \Eq{Eq_AppB_5} is lower than the IE bound \Eq{Eq_AppB_4}, or, in another term:\footnote{\Sec{Sec_TheBiggerIssue} explains the exceptions where these boundaries are not applicable.}
\begin{equation}\label{Eq_AppB_5_1}
Q_{TE} \leq 1-\prod_{i=1}^{m}(1-Q_i) \leq \sum_{i=1}^{m}Q_i
\end{equation} 
The second inequality in \Eq{Eq_AppB_5_1} can be proven by expanding $\prod_{i=1}^{m}(1-Q_i)$:
\begin{align}\label{Eq_AppB_6}
1 - \prod_{i=1}^{m}(1-Q_i) &=1-(1-\sum_{i=1}^{m}Q_i+\sum_{i<j}Q_iQ_j- ... +(-1)^{m}\prod_{i=1}^{m}Q_i) \nonumber\\
             &=\sum_{i=1}^{m}Q_i-(\sum_{i<j}Q_iQ_j- ... +(-1)^{m}\prod_{i=1}^{m}Q_i) \nonumber\\
             & \leq \sum_{i=1}^{m}Q_i
\end{align} 
For the first inequality in \Eq{Eq_AppB_5}, we refer the readers to the original proof given for system reliability bounds in \cite{Ref_201} and \cite{Ref_206}.  

The gap between the IE formula \Eq{Eq_AppB_3} and the EP approximation \Eq{Eq_AppB_5} is due to the common basic events between MCSs. As mentioned earlier, computing the effect of all the possible combinations of common elements (i.e., using \Eq{Eq_AppB_3}) is a painstaking process. However, we can still improve the upper bound at almost no additional effort if we include the effect of one specific case: the basic events that are common between \emph{all} MCSs. Let $E_{com}=\bigcap_{k=1}^{c}e_k$ represent the conjunction of all the basic events that are common between all MCSs and $\hat{E}_i$ the conjunction of the other events in $E_i$; thus: $E_i=E_{com}\cap\hat{E}_i$. Accordingly, $Q_{com}=\prod_{k=1}^{c}q_k$ is the unavailability measure for $E_{com}$, and $\hat{Q}_i$ the unavailability for $\hat{E}_i$; thus, $Q_i=Q_{com}\hat{Q}_i$. From \Eq{Eq_AppB_3}, we have:
\begin{align}\label{Eq_AppB_14}
Q_{TE} &= Pr\left(\bigcup_{i=1}^{m}E_i\right) \nonumber\\
             &=Q_{com}(\sum_{i=1}^{m}Pr(\hat{E}_i)-\sum_{i<j}Pr(\hat{E}_i \cap \hat{E}_j)+\sum_{i<j<k}Pr(\hat{E}_i \cap \hat{E}_j \cap \hat{E}_k)- ... \nonumber \\
             &+(-1)^{m+1}Pr(\bigcap_{i=1}^{m}\hat{E}_i)) \nonumber\\
             &=Q_{com} Pr\left(\bigcup_{i=1}^{m}\hat{E}_i\right) \nonumber\\
             &=Q_{com}\hat{Q}_{TE}
\end{align} 
From \Eq{Eq_AppB_5}, we know that $\hat{Q}_{TE} \leq 1-\prod_{i=1}^{m}(1-\hat{Q}_i)$. Therefore:
\begin{equation}\label{Eq_AppB_15}
Q_{TE}=Q_{com}\hat{Q}_{TE} \leq Q_{com}(1-\prod_{i=1}^{m}(1-\hat{Q}_i)) = (\prod_{k=1}^{c}q_k)(1-\prod_{i=1}^{m}(1-\hat{Q}_i)) 
\end{equation}

Relation \Eq{Eq_AppB_15} could also be derived from the original EP approximation \Eq{Eq_AppB_5} if we only substituted $Q_i$ by $Q_{com}\hat{Q}_i$ and rewrote \Eq{Eq_AppB_5}. In fact, some FTA tools refer to \Eq{Eq_AppB_15} as the EP approximation \cite{Ref_289,Ref_290,Ref_291}.

~

In summary, the following three upper bounds can be used for approximating the unavailability of a top event: 
\begin{equation}\label{Eq_AppB_16}
Q_{TE} \leq (\prod_{k=1}^{c}q_k)(1-\prod_{i=1}^{m}(1-\hat{Q}_i)) \leq 1-\prod_{i=1}^{m}(1-Q_i) \leq \sum_{i=1}^{m}Q_i
\end{equation}

A top event occurs when any of its MCSs occur while the other MCSs are not in a failed state. For $W_{TE}$ being the overall frequency of occurrence of the top event, we will have:
\begin{equation}\label{Eq_AppB_17}
W_{TE}=\sum_{i=1}^{m}W_i\prod_{\substack{j=1 \\  j\neq i}}^{m}(1-Q_j)=(\prod_{j=1}^{m}(1-Q_j))\sum_{i=1}^{m}(W_i/(1-Q_i))
\end{equation}

It is important to note that $W_{TE}$ and $Q_{TE}$ are not directly convertible because they are approximated based on different simplification assumptions. As an example, \Eq{Eq_AppB_17} does not consider the effect of common elements as does \Eq{Eq_AppB_14}.

FMR uses \Eq{Eq_MCSQ} and \Eq{Eq_MCSW} to calculate the likelihood of individual failure scenarios. For the collective SIF output failure, we use the lowest upper bound in \Eq{Eq_AppB_16} as an estimation of probability and \Eq{Eq_AppB_17} as the frequency measure.

\subsection{Example}

We earlier modeled the basic events in \Sec{Sec_BE_Example} and calculated the MCSs in \Sec{Sec_MCS_Example} for the fault tree shown in \Fig{Fig_FT_Ex1}. We now estimate the collective unavailabilities and frequencies. 

By using \Eq{Eq_MCSQ} and \Eq{Eq_MCSW}, we can calculate $Q_i$ and $W_i$ for individual MCSs:\footnote{All $W_i$s are in $\V{ph}$ unit (i.e., per hour).} 
\[
\begin{array}{lll}
E_1: S1 \wedge S3 ~~&~~ Q_1=q_S q_S=\V{7.67E-7}~~&~~ W_1=Q_1(w_S/q_S+w_S/q_S)=\V{8.75E-11}\\
E_2: S1 \wedge R1 ~~&~~ Q_2=q_S q_R=\V{4.20E-10}~~&~~ W_2=Q_2(w_S/q_S+w_R/q_R)=\V{5.26E-11}\\
E_3: S2 \wedge S3 ~~&~~ Q_3=q_S q_S=\V{7.67E-7}~~&~~ W_3=Q_3(w_S/q_S+w_S/q_S)=\V{8.75E-11}\\
E_4: S2 \wedge R1 ~~&~~ Q_4=q_S q_R=\V{4.20E-10}~~&~~ W_4=Q_4(w_S/q_S+w_R/q_R)=\V{5.26E-11}\\
E_5: L1 \wedge S3 ~~&~~ Q_5=q_L q_S=\V{4.90E-7}~~&~~ W_5=Q_5(w_L/q_L+w_S/q_S)=\V{2.80E-11}\\
E_6: L1 \wedge R1 ~~&~~ Q_6=q_L q_R=\V{2.69E-10}~~&~~ W_6=Q_6(w_L/q_L+w_R/q_R)=\V{3.36E-11}\\
E_7: V1 \wedge V2 ~~&~~ Q_7=q_V q_V=\V{4.29E-5}~~~&~~ W_7=Q_7(w_V/q_V+w_V/q_V)=\V{1.95E-8}\\
\end{array}
\]
For the top event $TE$, by using \Eq{Eq_AppB_17} and the lowest upper bound in \Eq{Eq_AppB_16}:
\begin{align}
& Q_{TE}=1-\prod_{i=1}^{7}(1-Q_i)=\V{4.497E-5} \nonumber \\
& W_{TE}=(1-Q_{TE})\sum_{i=1}^{7}(W_i/(1-Q_i))=\V{1.987E-8 ph} \nonumber 
\end{align}

\section{Beyond calculation}\label{Sec_BeyondCalc}

\subsection{Events and logic}

Fault trees model the occurrence of events. They do utilize logic symbols such as AND and OR. However, the information that flows through a fault tree is not logic; it is events. In logic, $C = A \wedge B$ means proposition $C$ is true if and only if both propositions $A$ and $B$ are true. In FTA, $C = A \wedge B$ only suggests that the simultaneous occurrence of events $A$ and $B$ leads to the occurrence of event $C$; and if so, that is only because the analyst defines it so. There are no logical relationships between the occurrences of $A$, $B$, and $C$.

Not only does the fault tree structure have no relation to abstract logic, but even as a tool, it is not used to compute Boolean values. The values that are used and displayed in FTA tools are frequencies and probabilities, not Boolean 0s and 1s. FTA is not designed to show whether the top event variable will become 0 or 1 if a particular pattern of 0s and 1s occurs at the basic event variables. Logical structures in fault trees only enable us to identify cut sets and calculate frequency and probability values -- as outlined earlier in Sections \ref{Sec_FTAMCSs} and \ref{Sec_QuantitativeFTA}.

\subsection{Frequency in AND and OR gates}

In calculating the combinations of events, an event should be considered an \emph{interval} in time, not a timeless moment. In the latter case, the probability of two events occurring at the same time will be (almost) zero. In real scenarios, too, it is commonly understood that while a hazardous event begins at a single moment in time, it does not end at that same moment. A fire break out, for instance, is not known only as the single moment when the fire breaks out, but also includes the time interval during which the fire is ongoing. 

With this view of events, let us revisit the role of AND and OR gates in FTA. Consider two independent events $A$ and $B$, with their corresponding time intervals $T_A$ and $T_B$. An AND gate expressed by $C = A \wedge B$ means either $A$ starts within the period $T_B$ after the moment $B$ starts, or $B$ starts within the period $T_A$ after the moment $A$ starts. That is the only way we can say $A$ and $B$ occur at the same time, because an exact coincidence between two independent events is not practical.

Let the frequency and probability of occurrence of $A$ be $w_A$ and $q_A$, and those of $B$ be $w_B$ and $q_B$. The frequency of $A$ occurring while $B$ has already occurred will be $w_A \cdot q_B$, and the frequency of $B$ occurring while $A$ has already occurred will be $w_B \cdot q_A$. Therefore, the frequency of $A \wedge B$ will be:
\begin{align}
w_C = w_A \cdot q_B + w_B \cdot q_A \label{Eq_AND1}
\end{align}
which matches up with \Eq{Eq_MCSW}, as we formulated earlier. 

The probability of individual events $A$ and $B$ (i.e., $q_A$ and $q_B$) can be calculated by the time intervals of the individual events: $q_A = w_A \cdot T_A$ and $q_B = w_B \cdot T_B$. Therefore, \Eq{Eq_AND1} can also be rewritten as:
\begin{align}
w_C = w_A \cdot w_B (T_A + T_B) \label{Eq_AND2}
\end{align}

For a numerical example, consider two independent events with frequencies $w_A = w_B = \V{0.01 py}$. Let us consider a 5-hour time interval equally for both events; that is, $T_A = T_B = \V{5 h}$. Given that 1 year equals 8760 hours, the probability of being in the time interval of event $A$ is $q_A = w_A \cdot (T_A / \V{8760}) = \V{5.7E-6}$. Similarly, $q_B = \V{5.7E-6}$. Using \Eq{Eq_AND1}, we have: $w_C = \V{0.01} \times \V{5.7E-6} + \V{0.01} \times \V{5.7E-6} = \V{1.14E-7 py} $. If we use \Eq{Eq_AND2} and $w_A = w_B = \V{1.14E-6 ph}$, we have $w_C = \V{1.14E-6} \times \V{1.14E-6} \times (\V{5} + \V{5}) = \V{1.3E-11 ph}$, which is equal to $\V{1.14E-7 py}$.

~

OR gates are slightly different. Where $D = A \vee B$, event $D$ occurs if $A$ occurs while $B$ hasn't yet occurred, or if $B$ occurs while $A$ hasn't occurred yet. Therefore, the frequency of $A \vee B$ will be:
\begin{align}
w_D = w_A (1 - q_B) + w_B (1 - q_A) \label{Eq_ORFreq1}
\end{align}
which is in line with \Eq{Eq_AppB_17} presented earlier. 

\begin{figure}[!h]  
\begin{center}
\includegraphics[scale=0.6]{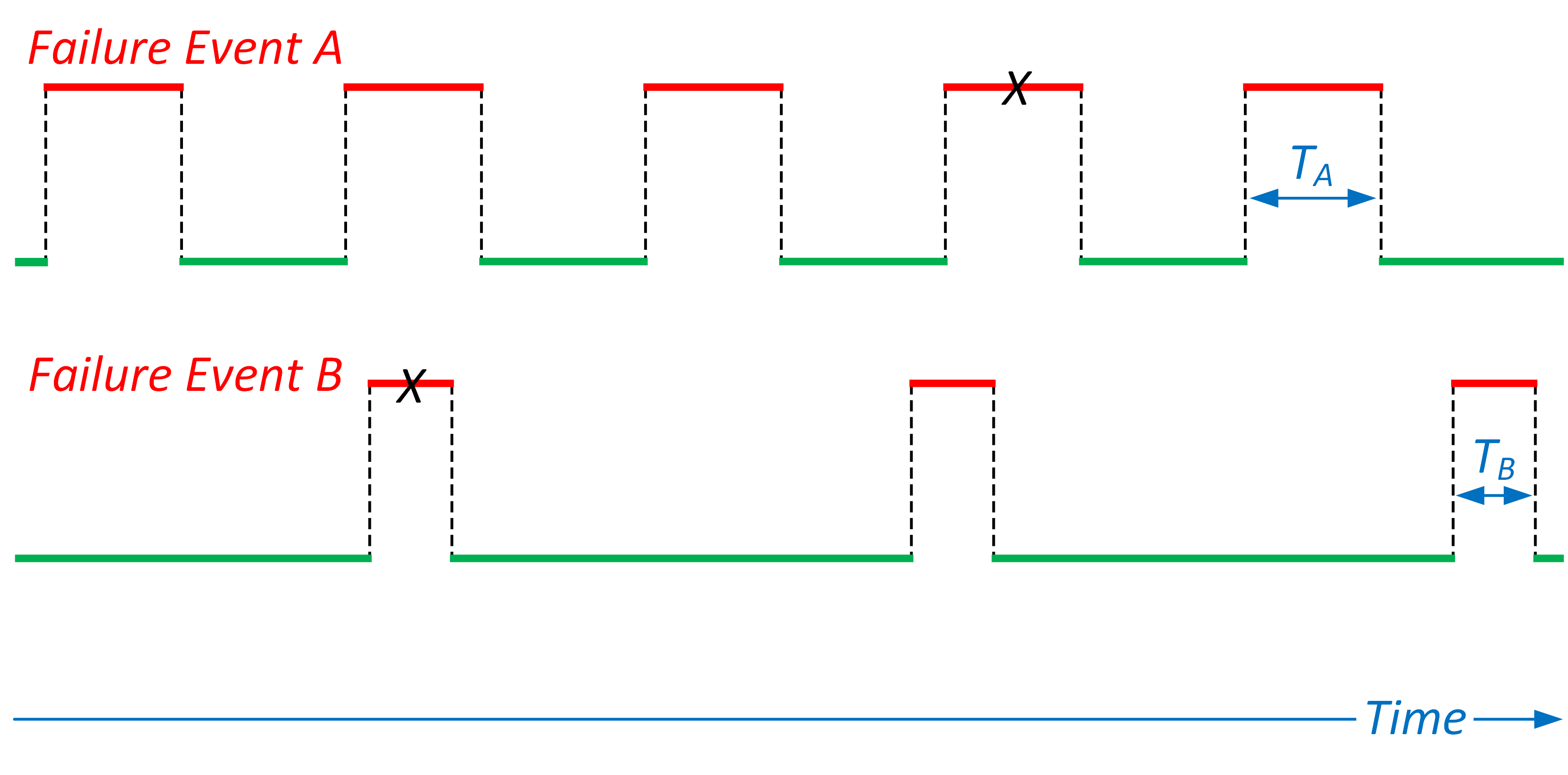}
\end{center}
\caption{Frequency calculation in OR gates}
\label{Fig_OR_Events}
\end{figure}

As \Eq{Eq_ORFreq1} suggests, in an OR gate, we only count the events that occur in a clear state (i.e., when neither failure event is present). This is because, as per the OR logic, the occurrence of one failure event already causes the occurrence of the overall failure; therefore, there is no need to count the second event. This is shown in \Fig{Fig_OR_Events}. This figure shows the occurrence of events $A$ and $B$ in the form of pulse trains, to represent both the frequency and the time interval of each event. The events that are marked with crosses are those that are not counted in the frequency calculation -- because they occur during the time interval of the other event.  

Display \Eq{Eq_ORFreq1} can also be written as follows:
\begin{align}
w_D =&\ w_A + w_B - (w_A \cdot q_B + w_B \cdot q_A)  \nonumber \\
=&\ w_A + w_B - w_{A \wedge B} \label{Eq_ORFreq2} 
\end{align}

Here, \Eq{Eq_ORFreq2} exactly represents what is depicted in \Fig{Fig_OR_Events}: the sum of both frequencies, subtracted by the concurrences of the two.

Recall the example we used earlier, where we had $w_A = w_B = \V{0.01 py}$ and $q_A = q_B = \V{5.7E-6}$. Using \Eq{Eq_ORFreq1}, the frequency of the OR combination will be $w_D = \V{2} \times \V{0.01} \times (\V{1} - \V{5.7E-6}) = \V{1.999988E-2 py}$. We could also use the $w_{A \wedge B} = \V{1.14E-7}$ calculated in the AND example and use \Eq{Eq_ORFreq2} to calculate $w_D = w_A + w_B - w_{A \wedge B} = \V{0.01} + \V{0.01} - \V{1.14E-7}$, which yields the same result.  

\subsection{NOT, a problem}

One particular case where the distinction between logic and FTA structure becomes clearly apparent is the NOT gate. In logic, NOT is an ordinary expression: in $A = \neg B$, $A$ is true if $B$ is false, and $A$ is false if $B$ is true. However, the use of NOT gates in FTA has long been a subject of debate \cite{Ref_392}. The original FTA methodology did not even include NOT gates \cite{Ref_389,Ref_387}. In some early resources, NOT was referred to as an \emph{extension} to FTA \cite{Ref_388,Ref_391}. 

The key issue with NOT is that it negates a failure to a \emph{success}; and including success in a model that is designed to analyze failures can introduce challenges. From a conceptual perspective, the key questions one may ask are: Why should the success of a component contribute to the failure of the system at all? Does that not indicate a problem in the system design? 

In reality, NOT does \emph{not} always indicate the success of a component. Some models use NOT gates simply to filter out the conditions in which the occurrence of failure is not enabled. In $D = (A \wedge B) \vee (\neg A \wedge C)$, the negation of $A$ simply means that $C$ is only relevant when $A$ is not present, whereas $B$ is only relevant when it occurs together with $A$. 

For a more practical scenario, consider an over-speed monitoring system installed on an industrial gas turbine that only runs for a few hours per day. Reliable over-speed monitoring systems are of significant importance. They monitor the turbine speed and activate an emergency shutdown if the speed is detected to be beyond critical thresholds. However, the criticality of failure of this system has different consequences depending on the operational state of the turbine. When the turbine is in operation, failure of the over-speed system can result in catastrophic destruction and casualties. However, failure of the system during the idle state may only cause minor operational delays. In such a case, the analyst may need to use NOT gates to filter out the non-critical failures.        

It is not always practical to avoid NOT gates in FTA models. However, one should be aware of their implications, as they can impact the outcome of the analysis in both qualitative and quantitative results.

One particularly interesting question that can be asked in relation to NOT gates concerns the frequency of an event: Where $A = \neg B$ and the frequency of event $B$ is $w_B$, what is the frequency of $A$? It is hard to offer a practical answer to this question, perhaps because $A$ is not a real event, but rather the \emph{absence} of one. However, if we have to put a number on $w_A$, we may simply say $w_A = w_B$, because the ``occurrence'' of $B$ can only happen if it hasn't yet occurred (i.e., it is not yet present). Therefore, for every occurrence of $B$, there would be one non-occurrence of it too, which means $w_A = w_B$.

\subsection{The bigger issue}\label{Sec_TheBiggerIssue}

The negation challenge is not limited to NOT gates. Other logic gates, such as NAND, NOR, and XOR, impose similar complications. This is debated in the broader context of ``non-coherence'' in Boolean structures, of which fault trees are one group. 

In simple terms, a coherent fault tree is one in which every basic event contributes to the overall system failure. Conversely, a non-coherent fault tree is one where some basic events move the system further away from the overall failure. A quick test to identify non-coherence is the use of negating gates (e.g., NOT). However, it should be noted that this is not a universal rule \cite{Ref_393}.

For a mathematical description, consider a set of components where each component can be in a binary state of success ($\V{1}$) or failure ($\V{0}$) at any given time. Let $x_i$ represent the state of the $i$th component and $\Bold{x} = (x_1, x_2, \dots, x_n)$ the state of the system. Let $\Phi(\Bold{x})$ be a Boolean expression over $\Bold{x}$, which represents the structure of a fault tree. We say $\Phi(\Bold{x})$ is ``coherent'' if the following three conditions are met:
\begin{itemize}[label=$\bullet$, itemsep=0.8em, left=2em]
    \item $\Phi(\V{1}_i, \Bold{x}) \neq \Phi(\V{0}_i, \Bold{x})$ ~for $1 \leq i \leq n$
    \item $\Phi(\Bold{\V{1}}) = \V{1}$, and $\Phi(\Bold{\V{0}}) = \V{0}$
    \item $\Phi(\Bold{x}) \geq \Phi(\Bold{y})$ ~if $\Bold{x} \geq \Bold{y}$
\end{itemize}

The first condition is known as component ``relevance,'' indicating that a change in the state of the component does affect the structure. The second condition implies that the system is in a working state when all components are in a working state, and in a failed state when all components are in a failed state. The third condition suggests that a coherent structure is a ``monotonic'' function, meaning that an additional component failure causes further degradation at the system level, and restoring any component to working condition reduces the level of degradation. Elaborate definitions and formulations of coherence can be found in various sources, including \cite{Ref_201,Ref_206,Ref_207,Ref_204}.

The concept of coherence is an important one in FTA. Recall the boundary conditions displayed in \Eq{Eq_AppB_5_1}. A critical presumption in those inequalities is the coherence of the structure. The reliability of a non-coherent system is not a monotonic measure and does not follow \Eq{Eq_AppB_5_1}.   

Non-coherent fault trees may not be as common as coherent ones; however, they do exist and are sometimes inevitable. We mentioned an example earlier -- related to gas turbine over-speed monitoring systems. More examples can be found in other papers, including, for instance, \cite{Ref_398}. What is more important is that the analyst is aware of how their FTA tool actually works and how it processes non-coherent fault trees.

\subsection{The ``right'' tool}

Depending on the tool and the algorithms it utilizes, the results of the analysis may differ -- both qualitatively and quantitatively. Some tools, such as Isograph's FaultTree+ \cite{Ref_289}, use basic Boolean logic rules to solve FTAs. Others, such as ITEM Toolkit \cite{Ref_291} and Astra-3 \cite{Ref_396}, use Binary Decision Diagrams (BDD) \cite{Ref_395}.  

Boolean logic simplification is a straightforward approach to simplifying complex fault trees into Minimal Cut Sets (MCSs). These rules are as follows:
\[
\begin{array}{ll}
A \vee (A \wedge B) = A ~~~~&~~~~ A \wedge (A \vee B) = A \\
A \wedge A = A  ~~~~&~~~~ A \vee A = A \\
\neg A \wedge A = \V{0} ~~~~&~~~~ \neg A \vee A = \V{1} \\
\neg (A \wedge B) = \neg A \vee \neg B ~~~~&~~~~ \neg (A \vee B) = \neg A \wedge \neg B \\
\end{array}
\]

Tools that use BDDs first convert the fault tree structure into a BDD, and then use it to compute the Prime Implicants (PIs), of which MCSs are a part. One algorithm for computing PIs is given in \cite{Ref_397}, and a practical example is shown in \cite{Ref_393}.  

The tools that calculate MCSs treat both failure and success events the same way. Those tool that produce PIs may produce unrealistic failure scenarios in addition to MCSs.

Some tools use coherent approximation, where the outputs of NOT gates are considered certain (non-probabilistic) events. Others give users the option to enable or disable the approximation option. 

Processing time may also be a consideration when choosing the FTA tool. Some tools use \Eq{Eq_AppB_3} to calculate $Q$ with high precision. In complex structures, the number of terms in \Eq{Eq_AppB_3} can be large, and the calculation loops must be cut off at some stage to ignore negligible contributions. This is a practical approach when the individual $q$ values are very small. However, in non-coherent structures, some of the $q$ values may be close to $\V{1}$, making the calculation very time-consuming before it converges to the desired precision.
    
There is no definitive answer to the question of ``right'' tool. Often, the choice depends on the needs and objectives of the analysis, as well as the budget and availability of skills. Nonetheless, it is important that the analyst understands how their tool actually works, so they can use it appropriately and achieve the objectives of their analyses.

\subsection{An example}

\begin{figure}[!h]  
\begin{center}
\includegraphics[scale=0.6]{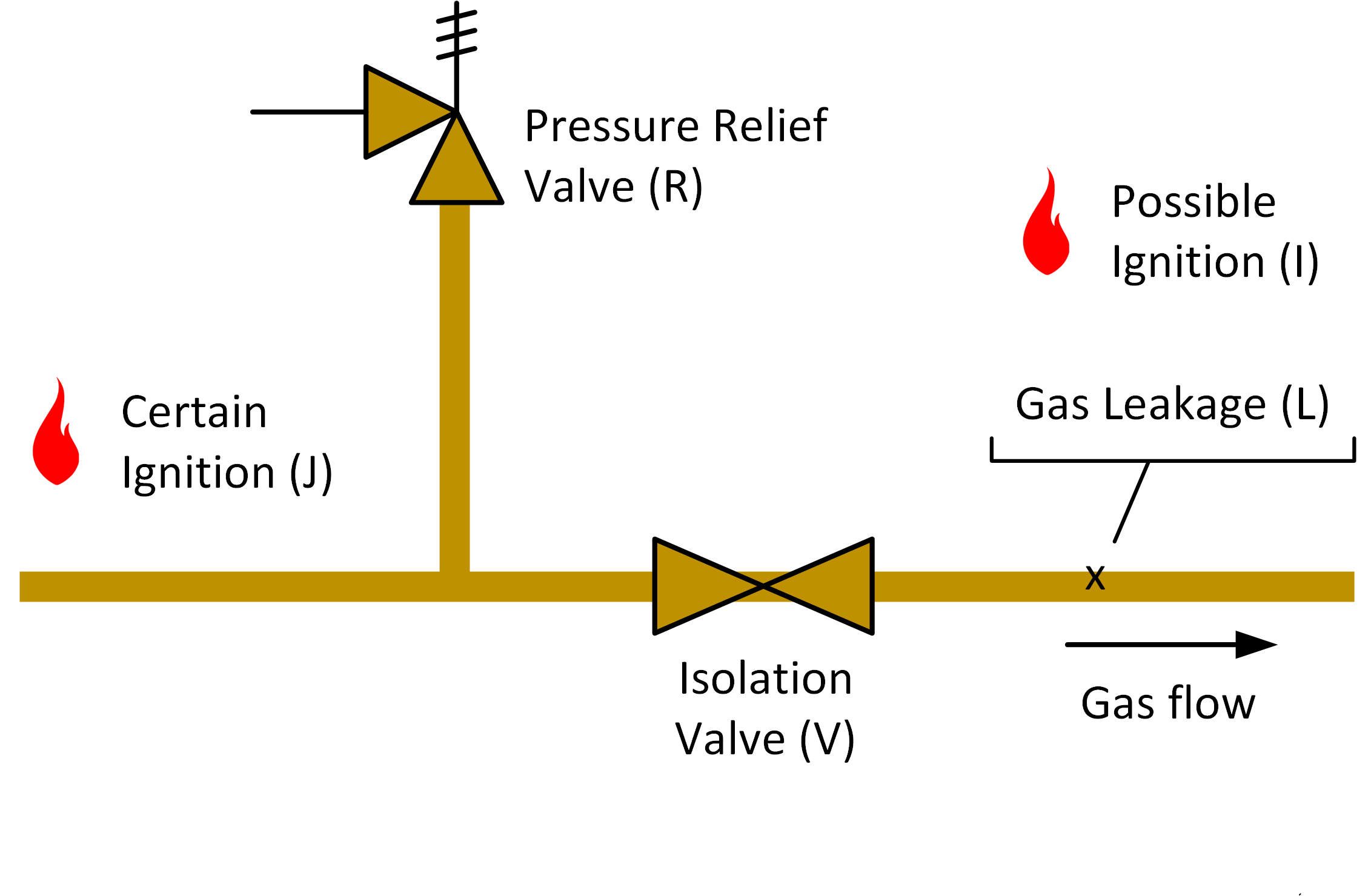}
\end{center}
\caption{Gas leakage scenario}
\label{Fig_LeakagePID}
\end{figure}

To better demonstrate the differences between coherent and non-coherent fault tree analyses, we present an example adopted from \cite{Ref_393}. Consider a gas leakage scenario as illustrated in \Fig{Fig_LeakagePID}. High-pressure gas flows through the pipe. If a leakage occurs in the section after the isolation valve (V), a gas detection system shuts the isolation valve. If the valve fails to close, the accumulation of flammable gas through the faulty pipe can lead to an explosion if the ignition source (I) ignites. If the valve does work -- and close -- excessive gas pressure will build up in the section of the pipe before the valve. To control the excessive pressure, a pressure relief valve (R) is installed, which automatically opens and vents the gas into a safe area in the atmosphere. If V succeeds and R fails, the pressure can cause a rupture in the pipe, which can in turn lead to an explosion in the area before the valve where the permanent ignition source (J) exists. 

\begin{figure}[!h]  
\begin{center}
\includegraphics[scale=0.6]{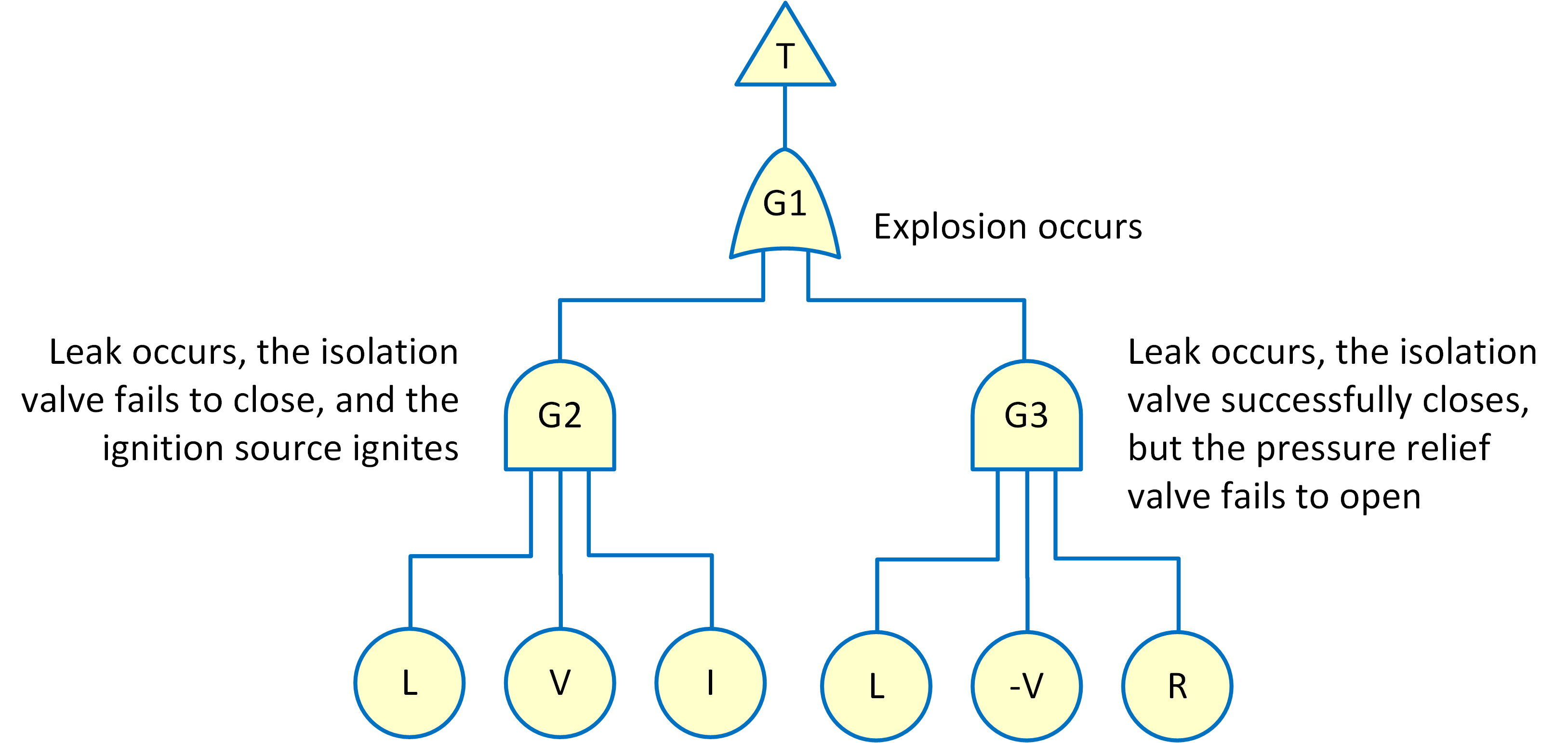}
\end{center}
\caption{FTA for the gas leakage scenario}
\label{Fig_LeakageFTA}
\end{figure}

\Fig{Fig_LeakageFTA} depicts the fault tree developed for the leakage scenario. The top event (i.e., explosion) occurs under the following two conditions:
\begin{itemize}[label=$\bullet$, itemsep=0.8em, left=2em]
\item A leak (L) occurs after the valve (V), the valve fails to close, and the ignition source (I) ignites.
\item A leak (L) occurs after the valve (V), the valve successfully closes, but the relief valve (R) fails to open, leading to a rupture in the area where the permanent ignition source (J) is.
\end{itemize}

Solving this fault tree by applying basic Boolean logic rules results in the following MCSs:
\begin{align}\label{Eq_LeakMCS}
\{L, V, I\} ~\text{and}~ \{L, \neg V, R\}
\end{align}

The issue is that an MCS is generally understood as a set of basic \emph{failure} events that can lead to the occurrence of the top failure event, and the second MCS in \Eq{Eq_LeakMCS} includes a \emph{success} element ($\neg V$). One may also question whether $\neg V$ can be considered an “event” in the first place, since it does not actually “occur.” What does occur is $V$ itself (valve failing to close), while $\neg V$ is present almost all the time.

One easy solution here is coherence approximation, where the negated element is simply removed. This relies on the assumption that basic failure events often have a very low probability of occurrence and, therefore, their complementary events are almost certain. Hence, we can use the approximation $(L \wedge \neg V \wedge R) \approx (L \wedge R)$, which reduces the previous MCSs to the following:
\begin{align}\label{Eq_LeakMCS1}
\{L, V, I\} ~\text{and}~ \{L, R\}
\end{align}

\begin{figure}[!h]  
\begin{center}
\includegraphics[scale=0.6]{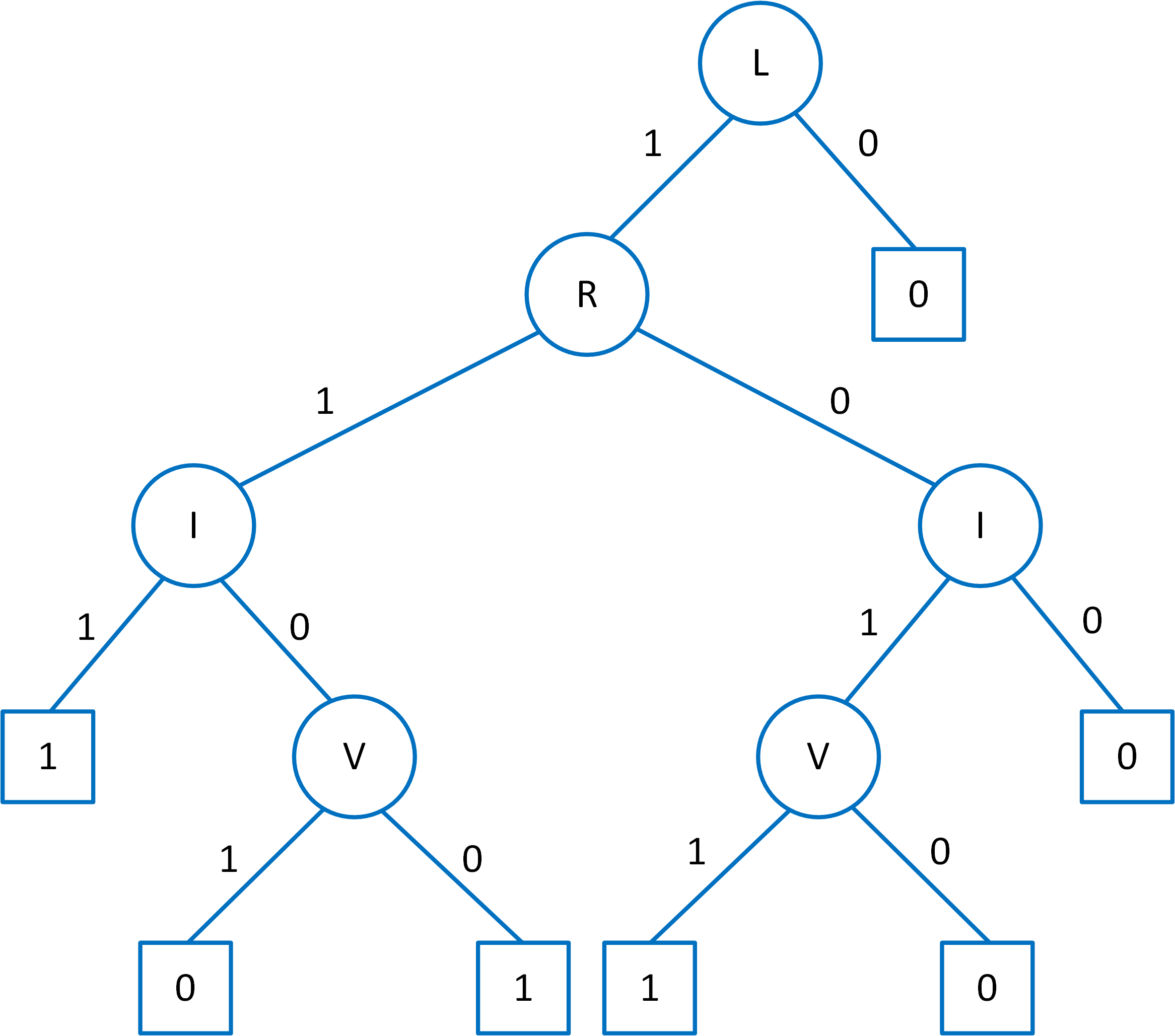}
\end{center}
\caption{BDD for the gas leakage scenario}
\label{Fig_LeakageBDD}
\end{figure}

Another approach is to use BDDs to generate the PIs instead of MCSs \cite{ref_394}. The corresponding BDD is shown in \Fig{Fig_LeakageBDD}, and the PIs derived from this BDD are as follows \cite{Ref_393}:
\begin{align}\label{Eq_LeakMCS2}
\{L, V, I\} ~\text{and}~ \{L, \neg V, R\} ~\text{and}~ \{L, R, I\}
\end{align}

The new element $\{L, R, I\}$ simply suggests that if a leakage occurs and the pressure relief valve (R) fails, and the ignition source (I) ignites, there will be an explosion irrespective of the success or failure of the valve (V). This is not wrong, but it is rather a purely mathematical implication rather than a practical incident scenario.

~

The difference between various approaches to non-coherent fault trees affects the quantitative analyses as well. For a quick comparison, let us consider fixed probability values for the failure events $L$, $I$, $V$, and $R$:
\begin{align}\label{Eq_LeakMCS3}
q_{_L}=q_{_I}=q_{_V}=q_{_R}=\V{0.05}; ~~~ q_{_{\neg V}}=\V{1}-q_{_V}=\V{0.95}
\end{align}

To calculate and compare the overall $Q_T$ of the top event, we use the four formulae shown in \Eq{Eq_AppB_16}. These formulae were derived in the previous section and used in \Eq{Eq_AppB_16} to compare boundary conditions for coherent structures. They are repeated here for ease of reference:
\begin{align}
Q_T &= Pr(\bigcup_{i=1}^{m}E_i) \label{Eq_QT1} \\
Q_T &= (\prod_{k=1}^{c}q_k) (1-\prod_{i=1}^{m}(1-\hat{Q}_i)) \label{Eq_QT2} \\
Q_T &= 1-\prod_{i=1}^{m}(1-Q_i) \label{Eq_QT3} 
\end{align}
\begin{align}
Q_T &= \sum_{i=1}^{m}Q_i \label{Eq_QT4}
\end{align}

The resultant $Q_T$ values of the top event are shown in \Table{Table_Cons}. The first column in \Table{Table_Cons} refers to the method used to derive the MCSs, or PIs; see Displays \Eq{Eq_LeakMCS}, \Eq{Eq_LeakMCS1}, and \Eq{Eq_LeakMCS2}. The other four columns show the values calculated using the different formulae above. For example, the third column uses \Eq{Eq_QT2} to calculate $Q_T$ for the MCSs shown in \Eq{Eq_LeakMCS} and \Eq{Eq_LeakMCS1} and for the PIs shown in \Eq{Eq_LeakMCS2}.

\begin{table}
\begin{center}
\begin{tabular}{|l|l|l|l|l|}
\hline \multicolumn{1}{|c|}{\TE{Method}}&\multicolumn{1}{|c|}{\TE{Using \Eq{Eq_QT1}}}&\multicolumn{1}{c|}{\TE{Using \Eq{Eq_QT2}}}&\multicolumn{1}{c|}{\TE{Using \Eq{Eq_QT3}}}&\multicolumn{1}{c|}{\TE{Using \Eq{Eq_QT4}}}\\
\specialrule{0.2em}{0.0em}{0.0em}
\TE{Boolean Logic as in \Eq{Eq_LeakMCS}} & ~\TE{\V{2.500000E-03}}$^*$~ & ~\TE{\V{2.494063E-03}}$^*$~ & ~\TE{\V{2.499703E-03}}~ & ~\TE{\V{2.500000E-03}}~ \\
\TE{Coherence Approx. as in \Eq{Eq_LeakMCS1}} & ~\TE{\V{2.618750E-03}}~ & ~\TE{\V{2.618750E-03}}~ & ~\TE{\V{2.624688E-03}}~ & ~\TE{\V{2.625000E-03}}~ \\
\TE{BDD as in \Eq{Eq_LeakMCS2}} & ~\TE{\V{2.500000E-03}}~ & ~\TE{\V{2.612827E-03}}~ & ~\TE{\V{2.624391E-03}}~ & ~\TE{\V{2.625000E-03}}~ \\
\hline
\end{tabular}
\end{center}
\caption{$Q_T$ values for different methods}
\label{Table_Cons}
\end{table}

From Table \ref{Table_Cons}, it is apparent that:
\begin{itemize}[label=$\bullet$, itemsep=0.8em, left=2em]
\item Depending on the method used, even when applying the same formula to calculate $Q_T$, the results may differ. See the values in each individual $Q_T$ column.
\item The inequality \Eq{Eq_AppB_16} may not hold for non-coherent structures. Consider the first two values in the first row -- marked with $^*$. In a coherent fault tree, and according to \Eq{Eq_AppB_16}, the first value would be smaller than the second one.
\end{itemize}

The quantitative differences in this example are negligible, and we use them here only to illustrate the point. However, this should not be generalized to all fault trees. Practical models often consist of large structures, where the impact may be considerable.

\subsection{The ``right'' fault tree}

One well-known example of Boolean expressions that is often discussed in the context of non-coherent fault trees is related to the ``Consensus Theorem.'' From logic, we know that:
\begin{align}\label{Eq_Consensus}
&(A \wedge B) \vee (\neg A \wedge C) =  (A \wedge B) \vee (\neg A \wedge C) \vee (B \wedge C)
\end{align}

In the context of FTA, this simply means that if either $A \wedge B$ or $\neg A \wedge C$ can cause an event, then $B \wedge C$ can cause the same event too -- irrespective of the status of $A$. We saw an example of such combinations in the gas leakage scenario, where $V$ and $\neg V$ were used. In that example, \Eq{Eq_LeakMCS} resembled the left-hand side of \Eq{Eq_Consensus}, and \Eq{Eq_LeakMCS2} the right-hand side. As discussed there, too, the two sides of \Eq{Eq_Consensus} can result in different analysis outcomes, even though they are logically equivalent. The question we would now like to address is: how can we choose which side of \Eq{Eq_Consensus} to use? In other words, is there a right side versus a wrong side, given that the two sides are logically equivalent?

Let us take a broader approach. We know from logic that every Boolean expression can be transformed into a Disjunctive Normal Form (DNF) \cite{Ref_226}. We also know that a fault tree is essentially a DNF -- an OR combination of a set of MCSs, each of which is an AND combination. The broader question we would like to answer here is: if there are two DNFs that are logically equivalent yet produce different qualitative and quantitative results in FTA, how do we decide which model is the “right” one to use?

Note that a primary assumption in this question is that the FTA models in question are \emph{correct}. If a model does not accurately represent the combinations of basic failures that can lead to the occurrence of the top event, the model should not be used. However, when it comes to choosing between two correct models -- like the two sides of \Eq{Eq_Consensus} -- the analyst may consider additional factors such as simplicity, practicality, and conservatism. 

Simpler models are often easier to verify and less prone to modeling errors. They also take relatively less time and effort to process and maintain. Hence, all other things being equal, they are usually the preferred choice.

With respect to practicality, the analyst may decide based on how useful a particular failure combination actually is. We explained earlier that the additional cut set $\{L, R, I\}$ in \Eq{Eq_LeakMCS2} is not incorrect, but it may not be a realistic scenario either. If leakage $L$ occurs, the status of valve $V$ will be known before we need to worry about the relief valve $R$. In other words, $\{L, R, I\}$ is not a real failure scenario, even though it is a theoretically valid combination of basic events that can logically trigger the top event. Readers may also refer to the traffic light example in \cite{Ref_392} for a similar case.

As for conservatism, one must recognize that failure analyses may be conducted with different objectives. Where risk and safety quantification is the goal, the analyst may decide to take a conservative approach and choose the model that indicates a higher failure probability. In such cases, the analyst may deliberately adopt a model that includes some unrealistic elements, in order to achieve a better safety margin.

To wrap up, let us highlight another interesting aspect of the Consensus Theorem. We explained at the beginning of this section that the logical structures used in FTA should not be confused with logic itself. The Consensus Theorem is yet another good example. While the two sides of \Eq{Eq_Consensus} are logically equivalent, they do not represent identical fault trees -- nor do they yield identical results. To choose which side is the ``right'' side, the analyst must consider engineering judgment in addition to mathematical correctness.

\section{Conclusion}

We presented a brief description of the FTA principles concerning qualitative and quantitative computations. We explained the concepts of modeling basic events, how MCSs are derived, and how frequency and probability values are calculated at the basic event level, MCS level, and top-event level. Furthermore, we explored the meaning of frequency in basic gates (AND, OR, and NOT), and examined some of the challenges related to negation, non-coherence, and considerations for choosing the most relevant model.

This paper did not aim to present a novel research idea. Rather, the goal was to cover some of the fundamental concepts in FTA computations, in order to help safety and reliability engineers gain a better understanding of their fault tree analyses. Such understanding can help improve the modeling of complex scenarios, which are often performed in software tools with limited visibility to the user.

\bibliographystyle{splncs04}
\bibliography{References}

\end{document}